\documentclass[12pt]{iopart}
\usepackage{rotating}
\usepackage{graphicx,pstricks,pst-plot,pst-node}
\usepackage{epstopdf}
\usepackage{latexsym,textcomp}
\usepackage{iopams}
\usepackage{amsfonts,amsthm,bm}

\begin{document}

\title{Prudent walks and polygons}

\author{John C. Dethridge, Timothy M. Garoni,  Anthony J. Guttmann and Iwan Jensen} 
\address{ARC Centre of Excellence for Mathematics and Statistics of Complex Systems, \\
Department of Mathematics and Statistics, 
The University of Melbourne, Victoria 3010, Australia}

\date{\today}

\ead{T.Garoni,T.Guttmann,I.Jensen@ms.unimelb.edu.au} 

\begin{abstract}
We have produced extended series for two-dimensional prudent polygons, based on a transfer matrix
algorithm of complexity O$(n^5),$ for a series of length $n.$ We have extended the definition to 
three dimensions and produced series expansions for both prudent walks and polygons in three dimensions.
For prudent polygons in two dimensions we find the growth constant to be smaller than that for the 
corresponding walks, and by considering three distinct classes of polygons, we find that the growth 
constant for polygons varies with class, while for walks it does not. We give the critical exponent 
for both walks and polygons. In the three-dimensional case we estimate the growth constant for both 
walks and polygons and also estimate the usual critical exponents $\gamma,$ $\nu$ and $\alpha.$
\end{abstract}

\submitto{\JPA}

\pacs{05.50.+q,05.70.Jk,02.10.Ox}

\maketitle

\section{Introduction}

A well-known long standing problem in combinatorics and statistical mechanics is
to find the generating function for self-avoiding polygons (or walks) on a two-dimensional
lattice, enumerated by perimeter. Recently, we have gained a greater understanding 
of the difficulty of this problem, as Rechnitzer \cite{AR03a} has {\em proved} that the 
(anisotropic) generating function for square lattice self-avoiding polygons is not 
differentiably finite \cite{RPS80a}, confirming a result that had been previously 
{\em conjectured} on numerical grounds \cite{GC01}.   That is to say, the generating function cannot 
be expressed as the solution of an ordinary differential equation with polynomial coefficients. There 
are many simplifications of the self-avoiding walk or polygon problem that are solvable \cite{BM96a},
but all the simpler models impose an effective directedness or equivalent constraint that reduces 
the problem, in essence, to a one-dimensional problem.

\begin{figure}[b]
  \begin{center}
    \includegraphics{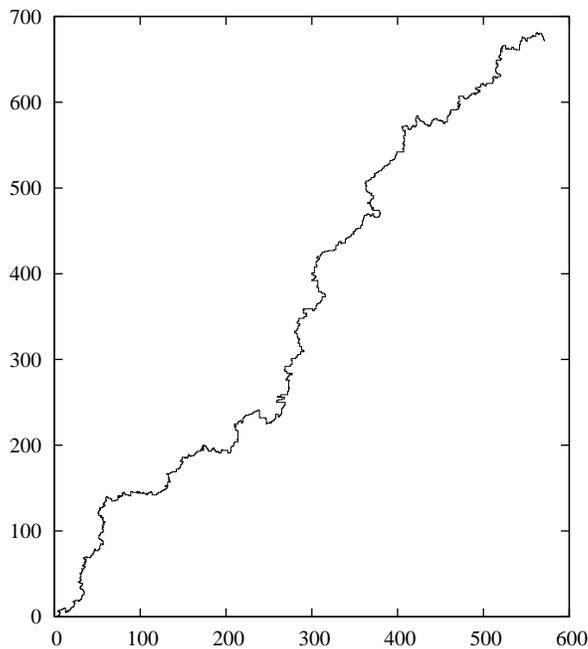}
  \end{center}
  \caption{\label{psaw} Typical prudent walk of $n=2000$ steps, generated via 
  Monte Carlo simulation using a pivot algorithm.}
\end{figure}

{\em Prudent} walks were introduced to the mathematics community  by Pr\'ea in an unpublished 
manuscript \cite{PP} and more recently reintroduced by Duchi \cite{D05}.
A prudent walk is a connected path on $ {\mathbb Z}^2$ such that, at each step, the extension 
of that step along its current trajectory will never intersect any previously occupied vertex. Such 
walks are clearly self-avoiding. We take the empty walk, given by the vertex $(0,0)$ to be a prudent walk.
Fig.~\ref{psaw} shows a typical prudent walk of $n=2000$ steps, generated via Monte Carlo simulation 
using a pivot algorithm. Note the roughly linear behaviour \--- it is believed, although unproven, 
that the mean-square end-to-end distance grows like $n^2$ for prudent walks, i.e. that $\nu=1$.

The {\it bounding box} of a prudent walk is the minimal rectangle containing the walk. The bounding box
may reduce to a line or even to a point in the case of the empty walk. One significant feature of 
two-dimensional prudent walks is that the end-point of a prudent walk is always on the boundary 
of the bounding box. Each step either lies along the boundary perimeter, or extends the bounding box. 
Note that this is not a bijection. There are walks such that each step lies on the perimeter of the bounding 
box that are not prudent. Such walks we call {\em perimeter walks}, and they will be the subject of a future 
publication~\cite{GG08}. Furthermore, if one extends the definition of prudent walks to three-dimensional 
walks, then {\it it is not true} that each step of the walk lies on the perimeter of the bounding box. Again, one 
can define three-dimensional walks with the property that each step lies on the perimeter of the bounding 
box, and these too will be discussed in the aforementioned publication~\cite{GG08}.

Another feature of prudent walks that should be borne in mind is that they are, generally speaking, not 
reversible. If a path from the origin to the end-point defines a prudent walk, it is unlikely that the path from 
the end-point to the origin will also be a prudent walk. Ordinary SAW are of course reversible.

A related, but not identical, model was proposed more than twenty years ago in the physics 
literature \cite{TD87}, where it was named the {\em self-directed walk}. In \cite{TD87} the authors 
conducted a Monte Carlo study and found that $\nu = 1$, where $\langle R^2 \rangle_n \sim cn^{2\nu}$. 
Here $\langle R^2 \rangle_n$ is the mean square end-to-end distance of a walk of length $n$. They 
also sketched an argument that the critical exponent $\gamma,$  characterising the divergence of the 
walk generating function, $C(x) = \sum c_n x^n \sim A(1 - \mu x)^{-\gamma}$ should be exactly 1, 
corresponding to a simple pole singularity.  This model differs from prudent walks in that different 
probabilities are assigned to different walks, depending on the number of allowable choices that 
can be made at each step. For the problem of prudent walks, all realisations of $n$-step walks are 
taken to be equally likely.

\begin{figure}
 \begin{center}
 \includegraphics[width=15cm]{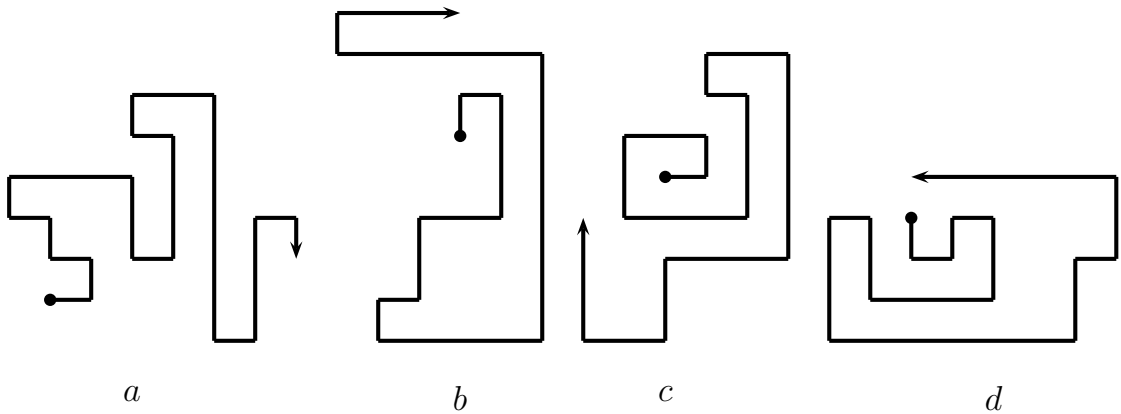}
    \end{center}
  \caption{\label{poly} Typical examples of: $a)$ a type 1 prudent walk, $b)$ a type 2 prudent walk, 
$c)$ a general prudent walk, and $d)$ a prudent polygon.}
\end{figure}

The problem proposed by Pr\'ea was subsequently revived by Duchi \cite{D05} who also studied two 
proper subsets,  called {\em prudent walks of the first type} and {\em prudent walks of 
the second type} (see figure~\ref{poly} for examples). 
Prudent walks of the first type are prudent walks in which it is forbidden 
for a west step to be followed by a south step, or a south step to be followed by a west step. 
Equivalently, prudent walks of the first type must end on the northern or eastern sides of the 
bounding box. Such walks are sometimes referred to as {\em 2-sided prudent walks. }  

Prudent walks of the second type are prudent walks in which it is forbidden for a west step to 
be followed by a south step when the walk visits the top of its bounding box and a west step 
followed by a north step when the walk visits the bottom of its bounding box. Equivalently, 
prudent walks of the second type must end on the northern, eastern or southern sides of their 
bounding box. Such walks are sometimes referred to as {\em 3-sided prudent walks. } Duchi found 
the solution for prudent walks of the first type, and gave functional 
equations for the generating function of (unrestricted) prudent walks. More recently the problem 
has been revisited by Bousquet-M\'elou~\cite{BM08}, who gave a systematic treatment of all three 
types, and in particular gave the solution for the generating function for prudent walks of the 
second type.

Results for prudent walks (unrestricted) can be found in \cite{DG08}.

\section{Computer enumeration \label{sec:enum}}

In two dimensions, we use transfer-matrix algorithms to count the number of prudent walks and polygons.
Given a prudent walk, only a small amount of information about the walk is required in order
to determine how to extend the walk by a single step so as to form a longer prudent walk.  
This information is called a \emph{configuration}.  Any prudent walk of a given length corresponds
to a unique configuration and there is only a finite number of configurations.  Our algorithm 
progresses by computing the number of walks of length $m$ corresponding to each possible 
configuration and then adding a single step at a time to each configuration, that is adding a step
to each walk in the equivalence class of walks corresponding to a given configuration. This 
is repeated (starting from a walk of size 1) until a given maximal length  $n$ is reached.  The 
information that needs to be stored in a configuration depends on which sub-class of prudent 
walks or polygons we are counting.

Consider unrestricted prudent walks.  Any partial walk of say $m$ steps will be contained within
a  bounding box, which is the smallest rectangle into which we 
can fit the walk.  It is easy to show that the end-point of the walk must be on a side of this box.  
The next step must either move away from the box, making it larger, or along the current side, 
which is possible only if the walk has not already visited vertices lying in that direction.

The information needed to describe the configuration of prudent walks is the dimensions 
of the bounding box,  which side of the box the walk end-point is on, the location of the end-point 
on that side, and an integer representing the directions in which the walk is allowed to be extended,
i.e. whether there are sites of the walk along the side to the left of the end, to the right of the end, 
or neither (both is not possible.)

Without loss of generality, we can assume that the end-point of the walk is  on the top side of 
the bounding box, since direction is not important for unrestricted prudent walks.  We can also 
assume that the end of the walk is not farther from the left side than from the right side, since
the number of ways in which a walk can be completed is invariant under reflection.

For a given configuration, there are three possible steps a walk  can take:

\begin{itemize}
\item A step outward, away from the current side;
\item A step to the left along the current side, if there are no walk sites in that direction;
\item A step to the right along the current side, if there are no walk sites in that direction.
\end{itemize}

When we move from a side to a corner of the bounding box, we consider the end-point of the 
walk to have moved to the other side only if that side has been moved outward by the step.

The transfer-matrix algorithm proceeds by adding a single walk step at a time.  We keep
a list of the possible configurations and the number of walks in the equivalence class
of that configuration.  The algorithm steps through the list, and for each source configuration $s$, 
generates the new target configurations $t_j$ that can be obtained by adding a single step (there
are at most three new target configuration $t_j$ per source configuration). We then 
add the number of walks $c_s$ with the source configuration $s$ to the number of walks 
$c_{t_j}$ of each of the target configurations $t_j$.  The length of a walk  is equal to
the number of iterations of the algorithm  and the total number of walks of a given length 
is found by summing the number of walks $c_s$ in the list of source configurations.

For prudent polygons, the position of the end-point of the walk is not enough information to 
determine  if the walk can be extended to create a prudent polygon.  The position of 
the start-point of the walk is also required.  So we have to extend a configuration to include 
the coordinates of the start-point of the walk.

If the start of the walk is not on one of the sides of the bounding box, nor adjacent to a site
on the current bounding box side that is in a direction in which the walk can extend, then 
the walk cannot be extended to form a prudent polygon, and we do not need to calculate  
the number of walks with that configuration.  So if such a configuration is generated by 
the transfer matrix algorithm, it is discarded.

If we are enumerating polygons of size $n$, and for a given configuration of $m$-step 
walks we cannot reach a site adjacent to the end point in $n-m$ steps or less, then we can 
ignore that configuration.

When calculating the number of polygons of a given size, we sum the number of walks only
for those configurations $s$ where the end-point is adjacent to the start-point.

The number of configurations grows like $n^4$. The width $w$ of the bounding box can 
vary from 1 to $n$ while the length $l$ can vary from 1 to $n-w$. Some smaller boxes cannot
occur because for $w\times l < n$ the walk cannot fit within the box, but we shall ignore this
effect for simplicity. In a box of size $w\times l$ the end-point can be in any of the $w$ positions
on the top edge while the starting point can be in $3w+2l$ positions, so up to constant factors
(arising from the restrictions on the directions the walk may take etc.) we have that the number
of configuration must be proportional to
$$ \sum_{w=1}^n \sum_{l=1}^{n-w} w(3w+2l) < 
 \sum_{w=1}^n \sum_{l=1}^{n} w(3w+2l)  < 
 \sum_{w=1}^n (3w^2n +2wn^2) = {\rm O}(n^4)
$$
As indicated above we do not have to keep all configurations but can discard some because
they won't give a prudent polygon of size $\leq n$. This however does not help reduce
the asymptotic growth in the number of configurations which remains at ${\rm O}(n^4)$
and the computational complexity of the algorithm is thus  ${\rm O}(n^5)$.

In prudent walks and polygons of types 1 and 2, we cannot ignore the direction of the current 
edge in our configuration, since some steps  may be disallowed depending on their direction.  
So we cannot assume that the end-point is on the top edge, or arbitrarily reflect the configuration, 
since rotations and reflections of a configuration are not equivalent.  However, for type 1 walks 
and polygons we can reflect about the southwest-to-northeast axis; and for type 2 walks we 
can reflect about the east-west axis.

Apart from the precise information required in the configuration, the test for which steps are 
valid from a given configuration, and which configurations should be accumulated to give
the result, the algorithm to enumerate each of these six objects (type 1, type 2 and full walks and polygons) 
is identical.  So we produced 
one program to solve all of these problems.  The lists are stored in hash tables for efficiency.  
The number of walks for each configuration and the total number of walks or polygons are 
computed modulo a large prime number close to the computer's word size.  The computation is 
repeated for several primes and the final result is calculated by use of the Chinese remainder
theorem.  This is more efficient than performing the whole calculation using numbers larger 
than the computer's word size.

In the following sections we show that for prudent walks and polygons of types 1 and 2
the generating functions can be derived rigorously or found from
relatively short series. The exceptions are unrestricted prudent walks
and polygons, and in the former case the number of walks can be
calculated efficiently by iteration of a functional equation. So the
only case requiring serious computational effort is that of
unrestricted prudent polygons. We enumerated the number of prudent
polygons up to size 1004. The calculation was performed on the SGI
Altix cluster of the Australian Partnership for Advanced Computing
(APAC). This cluster has a total of 1920 1.6GHz Itanium2 processors.  

The algorithm was parallelised in a fairly straightforward manner with configurations distributed 
across processors using a basic hashing scheme.  As in the basic algorithm, we step through
the list of source configurations and generate all the new target configurations. For each target
we check, using our hashing scheme, on which processor the target should reside. If this is not 
the current processor the configuration and its count is stored in a temporary stack. At regular 
intervals we pause in the main calculation in order to distribute the configurations from the 
temporary stacks to their designated processors. We found experimentally that it was advantageous 
to do the `parallel' hashing using only the bounding box and starting-point information from the 
configuration. Also doing the updating of the walk counts is fairly cheap and for this reason we 
did several primes simultaneously in a single run. To reproduce the integer coefficients correctly 
up to 1004 steps required some 36 primes so in practice we did 4 runs with each run using 9 primes. 
Each run utilised 160 processors and took about 9 hours with about 1/3 of this time used in the 
communications part of the algorithm. In total we used some 5500 CPU hours.

\section{Prudent walks}

In this section we summarise the known results for prudent walks. More detail can be found 
in \cite{BM08, DG08}.  We denote the generating function of prudent walks of the first type by
 $$C^{(1)}(x)=\sum c^{(1)}_n x^n,$$ where $c^{(1)}_n$ is the number of $n$-step 
prudent  walks of the first type. Then~\cite{D05}
\begin{equation}
\label{type1}
\fl
C^{(1)}(x)=1+x\frac{(1-2x-x^2)(3+2x-3x^2)+(1-x)\sqrt{(1-x^4)(1-2x-x^2)}}{(1-2x-x^2)(1-2x-2x^2+2x^3)}.
\end{equation}
It is clear that the dominant singularity is a simple pole located at the real positive zero 
of the polynomial $1-2x-2x^2+2x^3,$ notably at $x=x_c=0.4030317168\ldots .$  Thus the critical 
exponent $\gamma=1,$ and the asymptotic form of the coefficients is
$$c_n^{(1)} = \lambda^{(1)}/x_c^n+{\rm o}(\rho^{-n}),$$ for any $\rho < \sqrt{2}-1,$ where 
$\lambda^{(1)}=\frac{x_c(3x_c-1)}{(3x_c+1)(5x_c-2)} \approx 2.5165\ldots .$

We denote the generating function of prudent walks of the second type by 
$$C^{(2)}(x)=\sum c^{(2)}_n x^n$$ 
where $c^{(2)}_n$ is the number of $n$-step prudent  walks of the second type.

The generating function, first found by Bousquet-M\'elou~\cite{BM08} is much more complicated than 
that for prudent walks of the first type. First, we define
\begin{equation}
\label{q}
q\equiv q(x)=\frac{1-x+x^2+x^3-\sqrt{(1-x^4)(1-2x-x^2)}}{2x}.
\end{equation}
Then
$$C^{(2)}(x)=\frac{1}{1-2x-x^2}\left(2x^2qT(x;1)+\frac{(1+x)(2-x-x^2q)}{1-xq}\right)-\frac{1}{1-x},$$
where
\begin{equation}
\label{T}
\fl
T(x)\equiv T(x;w)=\sum_{k \ge 0}(-1)^k\frac{\prod_{i=0}^{k-1}(\frac{x}{1-xq}-U(q^{i+1}))}{\prod_{i=0}^k(\frac{xq}{q-x}-U(q^{i}))}
\left(1+\frac{U(q^{k})-x}{x(1-xU(q^k))}+\frac{U(q^{k+1})-x}{x(1-xU(q^{k+1}))}\right),
\end{equation} and
$$U(w) \equiv U(x;w) = \frac{1-wx+x^2+wx^3-\sqrt{(1-x^2)(1+x-wx+wx^2)(1-x-wx-wx^2)}}{2x}.$$

In this case the asymptotics are much more difficult to establish. Bousquet-M\'elou~\cite{BM08} 
has confirmed that the dominant singularity is precisely as for type 1 prudent walks, that is 
to say,  a simple pole located at the real positive zero of the polynomial $1-2x-2x^2+2x^3,$ 
notably at $x=x_c^{(1)}=0.4030317168\ldots .$ The factor 
$$\prod_{i=0}^k(\frac{xq}{q-x}-U(q^{i}))$$ 
appearing in the denominator of (\ref{T}) gives rise to an infinite sequence of poles on 
the real axis, lying between $x_c$ and $\sqrt{2}-1,$ which are not canceled by zeros of 
the numerator. This accumulation of poles is enough to prove that the generating function 
cannot be  D-finite.
 
Prudent walks have no additional geometric restrictions, the only restriction being that they are prudent.
We denote the generating function of such walks as $$C(x)=\sum c_n x^n,$$ where $c_n$ is the number 
of $n$-step prudent  walks. Duchi gave two coupled equations which can be iterated to give the series 
coefficients of prudent walks in polynomial time.  Rechnitzer\footnote{Private communication}   pointed 
out that these equations can be combined into a single equation,
\begin{eqnarray}
\label{full}
&&\frac{1}{xw}H(u,v,w)=1+H(u,v,w)+H(u,v,w)+H(u,v,w)+\\ \nonumber
&&\frac{xu}{v-xu}(H(u,v,w)-H(u,xu,w))
+\frac{xv}{u-xv}(H(u,v,w)-H(v,xv,w)),
\end{eqnarray}
which can be iterated, and the generating function obtained by setting $u=v=w=1.$ A closed 
form solution for this problem has not been found. In earlier work~\cite{DG08} the first 
400 series coefficients were obtained and analysed, and it was conjectured that the critical 
point and critical exponent remain unchanged from those of prudent walks of type 1 and type 2. 
That is to say, type 1, type 2 and unrestricted prudent walks have the same critical point, 
and the same critical exponent $\gamma=1,$ corresponding to a simple pole.

The anisotropic generating function can be defined as follows: If $c_{m,n}$ denotes the 
number of prudent walks with $m$ horizontal steps and $n$ vertical steps, then the 
anisotropic generating function can be written
$$C(x,y) = \sum_{m,n} c_{m,n}x^m y^n = \sum_n R_n(x)y^n,$$ where $R_n(x) = \frac{M_n(x)}{N_n(x)}$
is the (rational \cite{RPSECv2}) generating function for prudent walks with $n$ vertical steps.

We~\cite{DG08}  calculated the first 10 generating functions $R_1(x),\ldots,R_{10}(x)$ and found a regular 
pattern in the denominators $N_n(x),$ with factors corresponding to  cyclotomic polynomials of steadily 
increasing degree. If this pattern persists, the generating function cannot be D-finite, as the pattern of 
cyclotomic polynomials of increasing degree implies a build-up of zeros on the unit circle in the complex 
$x$ plane, and such an accumulation is incompatible with D-finite functions. As noted above, 
type-2 prudent walks are  not D-finite, and the numerical evidence here allows us to conjecture that 
(anisotropic) unrestricted prudent walks are also not D-finite.

\section{Prudent polygons \label{sec:pols}}

Polygon analogues of these three classes of walks can be naturally defined as walks of the given 
class that end at a vertex adjacent to their starting vertex. The relevant generating functions are
$$P^{(1)}(x)=\sum p^{(1)}_n x^n,$$ 
where $p^{(1)}_n$ is the number of $2n$-step prudent polygons of the first type,  
$$P^{(2)}(x)=\sum p^{(2)}_n x^n,$$ 
where $p^{(2)}_n$ is the number of $2n$-step prudent polygons of the second type, and 
$$P(x)=\sum p_n x^n,$$ 
where $p_n$ is the number of $2n$-step prudent polygons. See Fig.~\ref{poly}
for an example of a prudent polygon.
We have generated extensive isotropic and anisotropic series expansions for
prudent polygons.

Just as  we did for prudent walks above, if we distinguish between steps in the $x$ and $y$ direction, 
and let $p_{m,n}$ denote the number of prudent  polygons with $2m$ horizontal steps and $2n$ vertical steps,
then the anisotropic generating function for polygons can be written
$$P(x,y) = \sum_{m,n} p_{m,n}x^m y^n = \sum_n H_n(x)y^n,$$ where $H_n(x) = \frac{R_n(x)}{S_n(x)}$
is the (rational \cite{RPSECv2}) generating function for prudent polygons with $2n$ vertical steps.

\subsection{Type-1, or 2-sided prudent polygons.}

We generated more than 100 terms of the series for type 1 polygons, as described in the previous section,
which was more than sufficient to identify the generating function.

For prudent walks of the first type, we found, experimentally, that the generating function satisfies
a second order linear ODE,
$$\sum_{i=0}^2 P_i(x)f^{(i)}(x)=0,$$ where
$$P_0(x)=(-18+72x-99x^2+42x^3-30x^4+32x^5-7x^6+2x^7),$$
$$P_1(x)=(1-x)(-12+78x-148x^2+97x^3-39x^4+47x^5-9x^6+4x^7),$$
$$P_2(x)=x (1-x)^2 (1-3x-x^2-x^3) (3-7x+2x^2-x^3),$$
which can be solved to yield
\begin{equation}
        \label{p1}
        \fl
     xP^{(1)}(x)=\sum p_n^{(1)}x^{n+1} = \frac{(1-3x+x^2+3x^3)}{(1-x)} -\sqrt{(1-x)(1-3x-x^2-x^3)}.
\end{equation}

This result has recently been 
proved by Schwerdtfeger~\cite{S08}, who showed it can be derived from the known result for the generating 
function of bar-graph polygons, as type-1 prudent polygons are essentially bar-graph polygons.
From (\ref{p1}) it can be seen that
$$p^{(1)}_n = const. \mu_1^{2n} \times n^{-3/2} (1+ {\rm O}(1/n)),$$ 
where $1/\mu_1^2$ is the smallest positive root of the polynomial $1-3x-x^2-x^3.$
The smallest root is at $x_c=0.29559774\ldots,$ so $\mu_1 = 1.8392867\ldots.$
This should be compared to $\mu = 2.4811943\ldots$ for type-1 prudent walks.  
Clearly, prudent polygons are exponentially rare among prudent walks, unlike the analogous situation for 
ordinary SAW, for which it is known that the growth constants of SAW and SAP are the same.

Using the conventional exponent notation, so that $P^{(1)}(x) \sim const. (1 - \mu_1^2 x)^{2 - \alpha},$
we see that $\alpha = 3/2.$ 
       
\subsection{Type-2, or 3-sided prudent polygons.}

For prudent polygons of the second type, we again generated long series, but were unable to find the 
generating function by numerical experimentation.  Given the complexity of the known generating function 
for type-2 prudent walks, this is perhaps not surprising. Nevertheless, we were able to obtain quite precise 
numerical results, allowing us to conjecture
$$P^{(2)}(x) \sim const. (1 - \mu_2^2 x)^{2 - \alpha},$$
where, again,  $\alpha = 3/2.$ However, for $\mu_2$ we find  $\mu_2=2.023896\ldots,$ which is greater 
than $\mu_1$ and shows that type-1 prudent polygons are exponentially rare among type-2 prudent polygons, 
which are in turn exponentially rare among type-2 prudent walks.

Very recently Schwertdfeger~\cite{S08} has obtained the exact generating function for these polygons, 
incidentally confirming our numerical conjectures. He finds the generating function to be

$$P^{(2)}(x) = 2\left(\frac{x^2}{1-x} + B(x,1) + R(x)\right)$$ where
$$B(x,u) = \frac{1-x-u(1+x)x-\sqrt{x^2(1-x)^2u^2-2x(1-x^2)u+(1-x)^2}}{2xu}$$ and
$$R(x)=\sum_{k\ge 0}L((xq^2)^k)\prod_{j=0}^{k-1} K((xq^2)^k),$$
where $$q=\frac{1+x^2-\sqrt{1-4x+2x^2+x^4}}{2x},$$
$$L(x)=\frac{(1+x^2-(1-2x+2x^2+x^4)q)(B(x,q)+x)}{1-x(1+x)q-(x(1-x-x^3)q+x^2)(B(x,q)+x)}$$ and

$$K(x)=\frac{(1-x)q-1-((1-x+x^2)q-1)(B(x,q)+x)}{1-x(1+x)q-(x(1-x-x^3)q+x^2)(B(x,q)+x)}.$$ 
Asymptotic analysis of this expression \cite{S08} shows that the dominant singularity is given by 
the real positive zero of $x^5+2x^2+3x-2,$  which occurs at $x=x_c=0.49409642\ldots = 1/2.0238964\ldots,$ 
and the singularity of the generating function is a square-root singularity, just as for type 1 
prudent polygons. Both the location of the singularity and its exponent confirm our earlier numerical work.

Turning to the anisotropic generating function, we find numerically that
$$H_n = \frac{x^{n+1}P_{n}(x)}{(1-x^2)^{n}},$$ for $n$ even, and
$$H_n = \frac{x^{n+2}P_{2n}(x)}{(1-x^2)^{n}}$$ for $n$ odd. This denominator behaviour is not 
inconsistent with a D-finite generating function, but is precisely of the form
 observed for type-2 prudent walks,  which are not D-finite. It is therefore not surprising that the
generating function for the {\em isotropic} type 2 polygon is not D-finite, though the functional form 
of the anisotropic generating function above gives no clue as to this fact. However, this result has 
been confirmed, in the isotropic case, by Schwertdfeger~\cite{S08}, who proved non D-finiteness.

\subsection{Unrestricted, or 4-sided prudent polygons.}

Again we resorted to a numerical study. The methods we used to analyse and estimate the 
asymptotic behaviour of the series have all been described in \cite{Gu89}. We found from
our analysis that the series expansion is quite badly behaved, with evidence
of very strong sub-dominant asymptotic behaviour that mask the dominant asymptotic behaviour 
until quite large values of $n$ are reached.  Our initial attempts at a differential approximant 
analysis were not as convincing as they usually are. This was not really surprising, as there 
is evidence of a large number of singularities on the real axis, just beyond the critical point. 
This is the known situation for type-2 polygons, where there is an infinite number of poles in a 
small range of the real axis, just beyond the critical point. It is unremarkable that unrestricted 
polygons behave similarly. Nevertheless, using 3rd order inhomogeneous  approximants, and a series 
of 125 coefficients, corresponding to a maximum perimeter of 250, we were able to estimate the 
critical point at $x_c = 1/\mu_P^2 = 0.2267 \pm 0.0004,$ or $\mu_P = 2.100 \pm 0.002.$ The 
singularities appeared to be double roots, and we could not get a consistent estimate of the exponent.

Accordingly, we based our analysis on ratio type methods, which generally are much more slowly 
convergent than differential approximant methods, but have the advantage that convergence, when 
it eventually does take place, is more evident.

Indeed, we had to generate some 500 terms in the series 
(corresponding to polygons with perimeter of up to 1000 steps), before we could get a reasonably clear 
picture of the asymptotics, and even then it is not as unequivocal as we would like.  To estimate the 
critical point, we used a range of extrapolation methods to extrapolate the ratio of successive terms. 
We first used Wynn's algorithm, which is known to be slowly convergent, but robust.  The first iterates 
of the ratios were monotonically increasing, beyond $4.3987.$ This sequence should converge to $\mu_P^2.$  
The second iterate was also steadily increasing, suggesting that $\mu_P^2 > 4.4038.$ 
Higher iterates were unstable.

We next used Brzezinski's $\theta$ algorithm, which is known to be rapidly convergent under optimal 
circumstances.  The first iterates of the ratios were monotonically increasing for 84 terms, then 
decreasing until 187 terms, then increasing again, beyond $4.4155.$ This sequence may continue to 
exhibit oscillatory behaviour, so we are reluctant to use it as anything other than a guide.  The 
second iterates started oscillating quite early in the sequence. Higher iterates were unstable.

We next used the Levin $u$-transform, which is also known to be rapidly convergent in ideal circumstances.  
The first iterates of the ratios were monotonically increasing for the first 122 ratios, then steadily 
decreasing, and seemingly approaching an asymptote at a value of 4.4157. The second iterates reached a 
minimum at 198 terms, then steadily increased, and also approached an asymptote around 4.4157, though 
this is of no significance, as the previous iterates were all around that value, so one would expect 
an iteration of an almost constant sequence to give that constant value. Higher iterates were unstable.

Finally, a Neville table gave monotone first and second iterates. The first iterates were steadily 
increasing, suggesting $\mu_P^2 > 4.4107,$ while the second iterates increased more slowly, and gave 
$\mu_P^2 > 4.4128.$ The third iterates reached a maximum at 116 terms, and then a minimum at 201 terms, 
and then steadily increased. If this monotone trend continues, we can conclude $\mu_P^2 > 4.4136.$
Combining all these methods, we estimate $\mu_P^2 = 4.415 \pm 0.001$ or $\mu_P = 2.1012 \pm  0.00025$ 
which encompasses all the results.

\begin{figure}[t]
\begin{center}
\includegraphics[width=15cm]{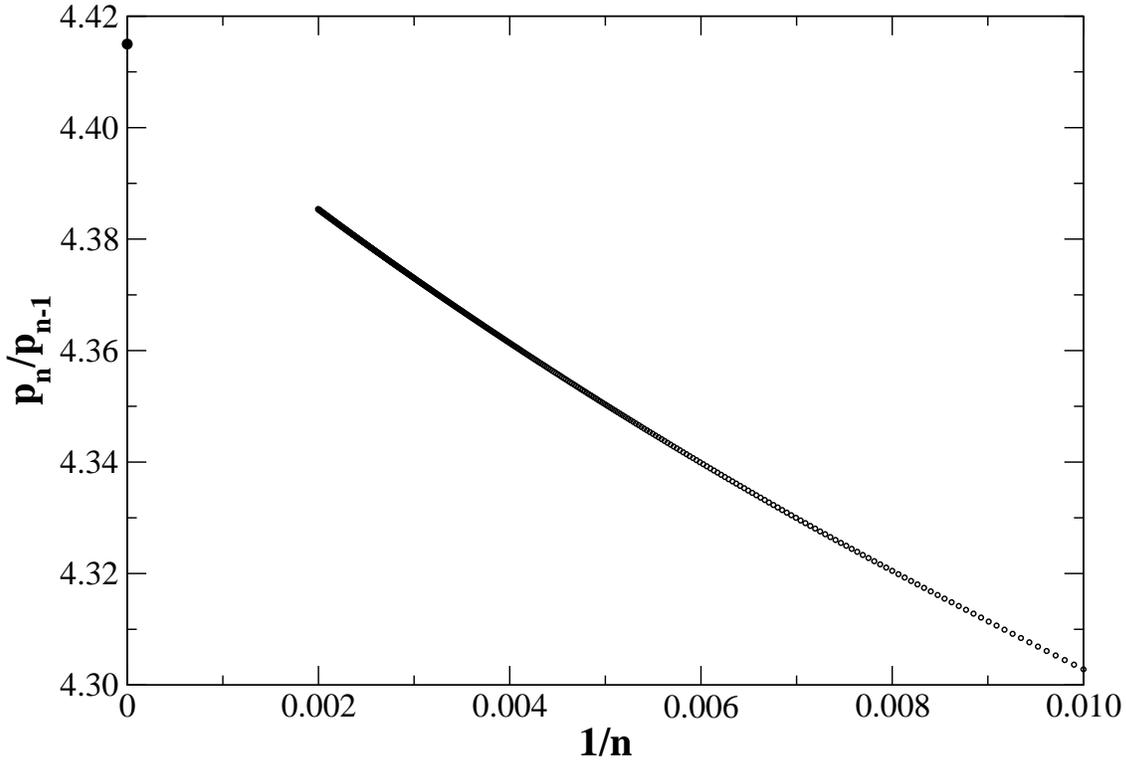}
\end{center}
\caption{\label{ratio} 
Plot of successive ratios of terms $p_n/p_{n-1}$ against $1/n$ for prudent polygons of half-perimeter $n$.
Our best estimate for the critical point $\mu_P^2=4.415$ is marked with a large dot on the $y$-axis.
}
\end{figure}

In figure \ref{ratio} we show a plot of the ratios of successive terms plotted against $1/n.$  Assuming 
the generating function behaves as $P(x) \sim const. (1 - \mu_P^2 x)^{2 - \alpha},$ then asymptotically, 
the ratios should approach  $\mu_P^2,$ with gradient  $\mu_P^2(\alpha-3).$ From the ratio plot, a limit 
around 4.415 is very plausible, though there is still a small amount of curvature in the locus of ratios. 
This is almost certainly due to the presence of exponentially small corrections due to one or more nearby 
singularities on the real axis. Extrapolating this locus to the estimated limit $\mu_P^2 \approx 4.415,$ 
we estimate the slope in the vicinity of the $y$-axis to be $-20,$ so that $\alpha \approx -1.5.$
We have also used biased differential approximants, with the critical point biased at $\mu_P^2 = 4.415,$ 
and while this signalled a confluent singularity at the critical point, it gave a consistent value for 
the exponent, $\alpha-2$ as $-3.5 \pm 0.1,$ so that again $\alpha \approx -1.5.$

 In summary, we find for the generating function,
$P(x) \sim const. (1 - \mu_P^2 x)^{2 - \alpha},$ where now $\alpha \approx -1.5.$ We do not quote error 
bars as this estimate of $\alpha$ is very sensitive to the estimate of the critical exponent used to bias 
the results. If we bias the critical point at  $\mu_P^2 = 4.416,$  a change of only 1 in the least 
significant digit, the exponent estimate from differential approximants changes to $-4.2 \pm 0.02.$ 
Thus our estimate of $\alpha$ can only be taken as indicative, rather than precise. What is clear however 
is that the singularity is unlikely to be of square-root type, as is the case for type 1 
and type 2 polygons, but of course there is no obvious reason why it should be.

For the anisotropic case, we write the generating function as
$$P(x,y) = \sum_{m,n} p_{m,n}x^m y^n = \sum_n H_n(x)y^n,$$ where the first few $H_n(x)$ are:\\
$$H_1(x)=\frac{1+x-x^2}{(1-x)^3}$$
$$H_2(x)=\frac{1+3x+2x^2+x^3}{(1-x)^5}$$
$$H_3(x)=\frac{1+5x+x^2-3x^3+x^5}{(1-x)^5(1+x)}$$
$$H_4(x)=\frac{1+7x+16x^2+18x^3+12x^4+4x^5+x^6}{(1-x)^7(1+x)}$$
$$H_5(x)=\frac{1+11x+19x^2+3x^3-12x^4+5x^5+6x^6-x^7-x^8}{(1-x)^7(1+x)^2}$$
$$H_6(x)=\frac{V_9(x)}{(1-x)^9(1+x)^2}$$
$$H_7(x)=\frac{V_{14}(x)}{(1-x)^9(1+x)^4(1+x+x^2)}$$
$$H_8(x)=\frac{V_{15}(x)}{(1-x)^{11}(1+x)^4(1+x+x^2)},$$
where $V_i(x)$ denotes a polynomial of degree $i.$
From the above, we see the relentless build-up of cyclotomic polynomials of
increasingly high order. As is well-known, if this pattern persists, the
underlying generating function cannot be D-finite. While this does not {\em a priori}
prove that the isotropic generating function is not D-finite, we know of no
combinatorial problem where this is the case. That is to say, where the isotropic
generating function {\em is} D-finite, while the anisotropic generating function is
not.

\section{Three-dimensional walks and polygons}
We have enumerated all prudent walks of up to $n=23$ steps on the
three-dimensional simple cubic lattice, using a simple backtracking
algorithm. As for the case of ordinary SAW, it is a far more difficult
task to perform enumerations in three dimensions than it is in two.
To take advantage of the inherent symmetry in the problem we only
explicitly counted walks whose first step was in the positive $x$
direction, whose first step out of this line (if any) was in the
positive $y$ direction, and whose first step out of this plane (if
any) was in the positive $z$ direction.  In addition, we {\em
  trivially parallelized} the backtracking algorithm: In three
dimensions there are sixteen 4-step prudent walks whose first two steps are
precisely East then North.  We independently enumerated the
completions to $n$-step walks of each of these prefixes.   Similarly
there are four 4-step prudent walks whose first three steps are East, East,
North. Again we independently enumerated the $n$-step completions of
each of these prefixes. Finally, the prudent walks whose first three
steps are East, East, East were all enumerated together in the same
computation.  Thus, for each value of $n$ we ran a total of 21
independent computations. The enumerations for $n=23$ took a total of
around 2377 hours (roughly 100 hours per independent computation).
The computations were performed on {\em tango}, a 95-node Linux cluster at
the Victorian Partnership for Advanced Computing (VPAC). Each node
consists of two AMD Barcelona 2.3GHz quad core processors.  In
principle, it is probably possible to obtain another term or two by
applying the two-step method~\cite{CLS07}, but we have
not pursued this here.

For each $1\le n\le 23$ we computed the number of prudent walks,
$c_n$, the number of prudent polygonal returns, $u_{n+1}$,  and the sum of squared
end-to-end distances $c_n\,\langle R^2\rangle_n$, summed over all
prudent walks.  The results of our enumerations are presented in
Table~\ref{3d enumeration data}.

\Table{\label{3d enumeration data} Exact enumeration data on the simple cubic lattice.
Here $c_n$ denotes the number of $n$-step prudent walks, $u_{n+1}$ the number of 
$(n+1)$-step prudent polygonal returns, and $\langle R^2\rangle_n$ the average 
squared end-to-end distance of $n$-step prudent walks.
}
\br
$n$ &          $c_n$  &   $c_n\,\langle R^2\rangle_n$  &        $u_{n+1}$  \\
\mr
  0 &                    1 &                      0 &                0 \\
  1  &                   6  &                     6  &                0  \\  
  2  &                  30  &                    72  &                0  \\  
  3  &                 150  &                   582  &               24  \\  
  4  &                 726  &                  4032  &                0  \\  
  5  &                3510  &                 25542  &              240  \\  
  6  &               16734  &                153048  &                0  \\  
  7  &               79518  &                881118  &             2544  \\  
  8  &              375246  &               4925616  &                0  \\  
  9  &             1766382  &              26909934  &            31800  \\  
 10  &             8278638  &             144356280  &                0  \\  
 11  &            38721366  &             762839334  &           435864  \\  
 12  &           180556206  &            3981064368  &                0  \\  
 13  &           840524742  &           20556000822  &          6323352  \\  
 14  &          3903866526  &          105173637672  &                0  \\  
 15  &         18106798830  &          533839505646  &         95647104  \\  
 16  &         83832778110  &         2690761186608  &                0  \\  
 17  &        387690560718  &        13478479905486  &       1493934516  \\  
 18  &       1790330065854  &        67142893855752  &                0  \\  
 19  &       8259528315558  &       332807521103670  &      23934001600  \\  
 20  &      38059497518214  &      1642214518277040  &                0  \\  
 21  &     175228328442174  &      8070246610372494  &     391427518152  \\  
 22  &     805959153119262  &     39511166688322248  &                0  \\  
 23  &    3704270575724550  &    192780251992208934  &    6511949001648  \\  
\br
 \endTable

We analysed the various series by the method of differential approximants, as well as 
variants of the ratio method, as used in the analysis of the polygon generating function in 
the previous section, (see \cite{Gu89} for a review of these standard methods). 
For the prudent walk generating function, for which we expect 
$$C(x) = \sum c_n x^n \sim const. (1 - x/x_c)^{-\gamma}$$ 
we found a singularity on the positive real axis at $x_c \approx 0.22265 \pm 0.0003$ 
with corresponding exponent $\gamma = 1.68 \pm 0.03.$ Biasing the value of the 
critical point at the central estimate, that is setting $x_c = 0.22265$ gives 
the corresponding biased estimate of $\gamma = 1.67$ from first order differential 
approximants, and $\gamma = 1.68$ from second-order differential approximants.  
For SAW, the analogous values are $x_c({\rm SAW}) \approx 0.2134907$ and 
$\gamma_{\rm SAW} \approx 1.1567,$ so prudent walks are exponentially rare among SAW, and the two models
have different critical exponents.

Additionally, for three-dimensional prudent walks we find that there is a singularity in the generating function $C(x)$ 
on the negative real axis that appears to be at, or just beyond, $x = -x_c.$ For the 
corresponding self-avoiding walk model, it is known that there is a singularity 
(the analogue of an anti-ferromagnetic singularity for a magnetic model) exactly at $x=-x_c({\rm SAW}),$ 
but the argument for the location of that singularity in the case of SAW does not translate 
to prudent walks. The corresponding exponent is, very approximately, the same magnitude as 
that of the physical singularity, but opposite in sign. That is to say, there appears to 
be a singularity of the form $$const.(1 + x/x^*)^{\gamma^*},$$  where $x^* \ge x_c$ and 
$\gamma^* \approx \gamma,$ as well as the physical singularity. For two-dimensional prudent 
walks, there is also evidence of a singularity on the negative real axis, but located 
considerably further away from the origin than the physical singularity. 
That is to say, $x^*(2d) > x_c(2d).$

To calculate the exponent $\nu$ characterising the mean square end-to-end distance, 
$\langle R^2 \rangle_n \sim const. n^{2\nu}$ we analysed the series for the sum of the 
squared end-to-end distances, which diverges at $x_c$ with exponent $\gamma + 2\nu.$ 
From a differential approximant analysis we found $x_c \approx 0.22265$ with 
$\gamma + 2\nu \approx 3.20,$ so that $\nu \approx 0.76.$ We also analysed the 
series $\langle R^2 \rangle_n$ directly, and obtained an estimate for $\nu$ consistent 
with that just quoted, but less precise.

The polygonal returns $u_n$ include a factor $2n$, corresponding to $n$ possible 
starting points and a factor of 2 as the path may be traversed clockwise or 
anticlockwise. We have analysed the series with coefficients $p_n=u_n/2n.$ 
We only have 11 coefficients, which is not really enough for any but the crudest analysis. 
From a differential approximant analysis, we find the generating function 
$P(z) = \sum_n p_nz^n$ is singular at $z = z_c \approx 0.0499 \pm 0.0006,$
with an exponent $2.3 \pm 0.5.$ That is to say, $P(z) \approx const. (1-z/z_c)^{2.3}.$ 
Note that  from our analysis of the corresponding walk series, $x_c^2 = 0.04957 \pm 0.00014.$ 
Thus it is entirely possible that $z_c = x_c^2$ for three-dimensional prudent polygons, 
just like the analogous situation for three-dimensional SAW, but unlike the situation 
for two-dimensional prudent walks and polygons, for which, as discussed above, 
prudent polygons are exponentially rare among prudent walks. If it is true that $z_c = x_c^2$ 
for three-dimensional prudent polygons, then we can carry out a biased analysis in which we 
fix $z_c$ at $x_c^2.$ In that case we find the exponent is a little higher, at $2.5 \pm 0.2.$ 
In terms of the usual notation for critical points, this exponent is $2 - \alpha,$ 
so that $\alpha = -0.5 \pm 0.2$

\section{Conclusion}
We have defined and analysed series for prudent polygons in both two and three dimensions. 
In two dimensions, we also discussed two subsets, which are exactly solvable. We found that 
two dimensional prudent polygons are exponentially rare among prudent walks, unlike the 
situation for two dimensional SAW. We gave numerical arguments in support of the conjecture 
that the generating function for two dimensional prudent polygons is not D-finite.

We also derived extensive series for three dimensional prudent walks and polygons. As far 
as we are aware, this problem has not been studied previously. We have given estimates 
of the critical point and critical exponents. In terms of the usual notation we found
that the growth constant for walks is $\mu \approx 4.491,$ with exponents $\gamma \approx 1.68,$ 
$\nu \approx 0.76,$ and $\alpha \approx -0.3$ based on an unbiased estimate of the 
critical point. For SAW $\mu_{\rm SAW} \approx 4.68404$, $\gamma \approx 1.1567,$  
$\nu \approx 0.5875$ and $\alpha \approx 0.2375.$ For SAW we have the hyper-scaling relation 
$d\nu = 2 - \alpha.$ While there is no {\it a priori} reason to expect this hyper-scaling 
relation to hold for three-dimensional prudent walks, as it is a combinatorial model rather than a 
statistical mechanical model derived from a Hamiltonian, we nevertheless note that 
$3\nu \approx 2.28$ while $2 - \alpha \approx 2.28.$ The uncertainties in our exponent 
estimates are too great to attach much significance to this approximate equality, 
except to flag it as a possibility.

In terms of physical significance, the model of prudent walks, while exponentially rare 
among SAW in both two and three dimensions, is nevertheless exponentially abundant 
compared to any other solved or numerically estimated model. Other variants of the model 
that relax the prudency constraint to some extent are likely to be have growth constants 
even closer to that of SAW, and these will be the subject of a future publication~\cite{GG08}.

\ack
 
We would like to thank Simone Rinaldi and Enrica Duchi for introducing us
to this problem, and Mireille Bousquet-M\'elou for several enlightening discussions, and access 
to her results prior to publication. Similarly, we thank Uwe Schwerdtfeger for provision of his
results prior to publication.
The calculations presented in this paper  were performed on the facilities of the
Australian Partnership for Advanced Computing (APAC) and the Victorian 
Partnership for Advanced Computing (VPAC). 
We gratefully acknowledge financial support from the Australian Research Council.


\section*{References}


\begin{thebibliography}{10}

\bibitem{AR03a} Rechnitzer A 2003 
Haruspicy and anisotropic generating functions 
{\em Adv.  Appl. Math.\/} {\bf 30} 228--257

\bibitem{RPS80a} Stanley R~P 1980 
Differentiably finite power series 
{\em European J. Combin.\/}  {\bf 1} 175--188

\bibitem{GC01}
Guttmann A~J and Conway A~R 2001 Square lattice self-avoiding walks and
  polygons {\em Ann. Comb.\/} {\bf 5} 319--345

\bibitem{BM96a} Bousquet-M\'elou M 1996 
A method for the enumeration of various classes of column-convex polygons {
\em Disc. Math.\/} {\bf 154} 1--25

\bibitem{PP} Pr\'ea~P 1997
Exterior self-avoiding walks on the square lattice. Unpublished manuscript.

\bibitem{D05} Duchi~E 2005
On some classes of prudent walks. In {\em FPSAC'05} Taormina, Italy, 2005.
    
\bibitem{GG08} Garoni T M and Guttmann A J 2008
Perimeter and quasi-prudent self-avoiding walks and polygons (in preparation).

\bibitem{TD87} Turban~L and Debierre~J.-M 1987
Self-directed walk: a Monte Carlo study in two dimensions 
{\em J. Phys. A: Math. Gen.\/} {\bf 20} 679--686
  
\bibitem{BM08} Bousquet-M\'elou M 2008
Families of prudent self-avoiding walks, arXiv hal-00276681
  
\bibitem{DG08}  Dethridge J C and Guttmann A. J 2008
  {\em Entropy} {\bf 10} 309-318
 
\bibitem{RPSECv2}  Stanley R P 1999
{\it Enumerative Combinatorics } vol. 2 (Cambridge: Cambridge University Press)
   
\bibitem{S08}  Schwertdfeger U 2008 
  Exact solution of two classes of prudent polygons, arXiv math.CO 0809.5232 

\bibitem{Gu89} Guttmann A J (1989) Asymptotic analysis of coefficients
in {\em Phase Transitions and Critical Phenomena}, 
eds C Domb and J L Lebowitz, (Academic: London)  {\bf 13} 1-234
  

\bibitem{CLS07} Clisby N, Liang R and Slade G 2007 
Self-avoiding walk enumeration via the lace expansion,
{\em J. Phys. A: Math. Gen.\/} {\bf 40} 10973--11017


\end{thebibliography}
\end{document}